\documentclass[conference]{IEEEtran}
\IEEEoverridecommandlockouts
\usepackage{cite}
\usepackage{amsmath,amssymb,amsfonts}
\usepackage{algorithmic}
\usepackage{graphicx}
\usepackage{textcomp}
\usepackage{ mathrsfs,dsfont }
\usepackage{xcolor}
\newtheorem{theorem}{Theorem}[section]
\newtheorem{lemma}[theorem]{Lemma}
\def\BibTeX{{\rm B\kern-.05em{\sc i\kern-.025em b}\kern-.08em
    T\kern-.1667em\lower.7ex\hbox{E}\kern-.125emX}}

\begin{document}
\setlength{\abovedisplayskip}{2pt}
\setlength{\belowdisplayskip}{2pt}
\title{A Routing and Link Scheduling Strategy for \\Smart Grid NAN Communications\\
\vspace{-0.15in}}
\author{\IEEEauthorblockN{Shuchismita Biswas and Virgilio A. Centeno}
\IEEEauthorblockA{Virginia Tech, Blacksburg, VA, USA}
\{suchi,virgilio\}@vt.edu
\vspace{-0.15in}}
\IEEEoverridecommandlockouts
\maketitle
\IEEEpubidadjcol
\vspace{-0.1in}
\begin{abstract}
As large scale deployment of smart devices in the power grid continues, research efforts need to increasingly focus on efficient communication of generated information. This paper describes a strategy for static routing and scheduling of messages in a multi-hop wireless Smart Grid Neighborhood Area Network (NAN) with multiple source nodes and a common set of destinations or gateways. The problem is formulated as a Mixed Integer Linear Program (MILP) and solved using commercial optimization solver CPLEX. Feasibility of the scheme is demonstrated using different network models, constraints, message injection rates, and initial conditions. It is shown that the proposed approach can be used to generate an optimal link schedule for collecting user-generated bids in a transactive energy market in the least possible time. It is also shown that the methodology is applicable to multiple destination nodes and that their location affects message delivery time. 
\end{abstract}

\begin{IEEEkeywords}
smart grid, NAN, wireless mesh networks, routing and scheduling, smart meters, transactive energy, MILP
\end{IEEEkeywords}
\vspace{-0.2in}
\section{Introduction}\label{sec:intro}
\begin{figure}[t]
    \centering
    \hspace*{.1in}
    \includegraphics[clip, trim=50 80 180 110, width=0.45\textwidth]{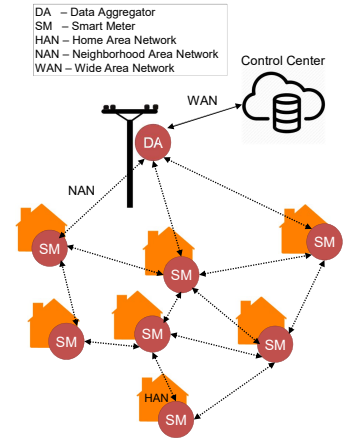}
    \caption{Smart grid communication architecture}
    \label{fig:comm_archi}
    \vspace{-0.28in}
\end{figure}

The smart grid (SG), essentially, is a cyber physical system where a communication layer is overlaid on the physical electrical grid. Individual components can now communicate with each other, paving the way for better control, monitoring, protection and informed decision-making. Information flows among the physical, sensor, communication and decision-making layers and any undesired operation may adversely impact the overall system performance \cite{4D}. 
The SG communication architecture may be divided into three parts- a) Home Area Network (HAN), b) Neighborhood Area Network (NAN) and c) Wide Area Network (WAN) \cite{NAN_survey}, as shown in figure \ref{fig:comm_archi}. Devices within a home communicate via technologies like Bluetooth and Zigbee, forming the HAN. NANs relay control, pricing and energy usage information among home devices and common access points or gateway nodes. These gateway nodes have better communication backhaul support, may be located at substations or utility poles, and  can communicate with a utility control center via the WAN.  WANs may employ fibre optic cables or cellular networks like WiMAX, GPRS and 3GPP. SG communications are bidirectional. 

Because NANs cover large and complex geographic expanses, their architecture needs to be secure, reliable, scalable, and cost-effective. They should be self-healing, i.e. the network needs to be capable of providing an acceptable level of Quality of Service (QoS) even if some links are out \cite{Qosaware}. 
Multihop Wireless Mesh Networks (WMN) are poised to be well-suited for NAN applications due to their flexibility, dynamic reconfiguration capabilities, cost-effectiveness and ability to cover large areas via relay nodes\cite{NAN_survey,wireless_economy}. Other related technologies like Broadband Power Line Communication (BPLC), Wireless Local Area Network (WLAN) and Wireless Sensor Networks (WSN) \cite{QoS_routing_WSN} may also be used to complement WMNs \cite{NAN_survey,rinaldi}. 
 However, before WMNs can be deployed on a large scale in NANs, communication protocols dictating how to collect, aggregate and disseminate information are needed. 
 
 There are several standardization efforts dealing with the issue of routing and packet scheduling in SG wireless mesh networks \cite{802_11s,SUN}, but they primarily focus on the Physical (PHY) and Media Access Control (MAC) layer, and upper layers like the network layer remain to be addressed. 
In \cite{QoS}, a routing algorithm for SG messages with multiple QoS requirements is proposed. The network is represented by a weighted graph, whose edge weights are vectors representing QoS variables. An auxiliary weight is then assigned to each edge and the shortest path from the source to destination is computed. The authors show that this is a $K$-approximation algorithm, where $K$ is the number of QoS constraints imposed. However,  waiting times at intermediate nodes is not considered while determining message paths. Multi-gateway routing is addressed in \cite{multigate} with a reserve path diversity scheme. All nodes have a tree table that defines the best path to a gateway. The source node starts transmitting on the best path according to its tree table and switches over to the second best route if a link breakage message is received. If both routes fail, then a timer-based on demand protocol is activated. The authors also propose a backpressure \cite{neely} based packet scheduling scheme for dynamic gateway selection. Here, the source node selects a neighboring node that minimizes the product of queue size and best path metric as its next hop. The selection procedure continues until one of the gateways is selected as the best neighbor. Other backpressure based scheduling strategies have also been proposed in literature \cite{VQ}, but they do not consider the fact that data rates in SG might be too low to effectively develop a backpressure gradient among neighboring nodes. 

This paper addresses these limitations and proposes a Mixed Integer Linear Program (MILP) based static routing and link scheduling strategy. Our approach minimizes the total time required to deliver all messages in a network to destination nodes. If all messages cannot be delivered within a stipulated time, then a schedule to minimize the number of undelivered messages is generated. The feasibility of this scheme is demonstrated using different network conditions and constraints. The proposed method is also used to generate an optimal schedule for collecting user generated bids in a transactive energy market in the least possible time.



The rest of the paper is organized as follows. Section \ref{sec:NAN} describes SG NAN communications and associated challenges in further detail. Section \ref{sec:problem} discusses the network model considered and formulation of the proposed MILP. Section \ref{sec:experiments} demonstrates the feasibility of our methodology using different use cases. Section \ref{sec:extension} discusses limitations of and possible extensions to our approach. Section \ref{sec:conclusion} concludes the paper.

\section{Neighborhood Area Networks}\label{sec:NAN}

\subsection{NAN Communication Standards}
In recent years, a number of standards related to SG communication networks have been introduced. Two of these include IEEE 802.15.4g \cite{SUN} and IEEE 802.11s\cite{802_11s}.

\subsubsection{IEEE 802.15.4g}IEEE 802.15.4g modifies the existing IEEE 802.15.4 standard to include PHY and MAC layer specifications for wireless Smart metering Utility Networks (SUN). SUNs are expected to connect a large number of mostly outdoor devices distributed over a large area with shared resources. Thus the primary network topology is peer-to-peer and some end-devices might need to send messages on a multi-hop path to reach a gateway. Communication among network devices is facilitated by a centralized network coordinator. SUNs operate in license exempt frequency bands (700 MHz to 1 GHz, 2.4 GHz) and have low data rates between 40 to 1000 kbps. They are also desired to coexist with other 802.11/802.15/802.16 networks that operate in the same frequency spectrum. 


\subsubsection{IEEE 802.11s} IEEE 802.11s  extends the MAC protocol of the existing IEEE 802.11 to self-configuring multi-hop WMNs. IEEE 802.11s mesh networks deploy a central gateway and multiple Mesh Access Points (MAP). End users can access MAPs in static or dynamic states and the MAPs send aggregated information to the gateways via multihop paths. The  default path selection protocol called Hybrid Wireless Mesh Protocol (HWMP) combines on-demand routing protocol with tree-based static routing protocol. Static protocols are used to maintain state while on-demand or reactive protocols
deal with changes in network topology.

An important difference between the 802.11s and 802.15.4g standards is that as the former operates on
the same PHY layer as the 802.11 family, high-speed data transmission is supported. The 802.15.4g, on the other hand is meant for low data rate utility networks. Both these standards deal with the PHY and MAC layers and are still being improved. Hence, to deal with the communication requirements in the future power grid, research on the upper layers of the protocol stack is required.  

\subsection{Data Traffic in NAN}
As mentioned before, SG NANs support different applications, including collecting billing information, distributing control commands, energy prices for demand response, and even firmware upgrades. This heterogeneous data, evidently, has different QoS requirements. Depending on their frequency, the SG data traffic maybe broadly classified into two classes. 
\subsubsection{Event-triggered} This class of traffic is generated due to an external event. Typically, they have low size (hence, low bandwidth requirement) and high priority (high reliability and low latency). Fault notifications, alarms, control commands etc belong to the event-triggered class of data traffic. On demand polling of information initiated by the control center (for applications like real time state estimation) also belong to this traffic category.
\subsubsection{Periodic} Periodic data are generated at pre-defined intervals. Their QoS requirements can be quite diverse. Message sizes may vary from medium to high and priority may be medium or low. Most AMI data are periodic in nature. Automated Meter Reading (AMR) messages have large size and low latency requirements ($>$ 2 hours). On the other hand, data meant for near real time applications may have lower packet size and delay tolerance. In \cite{PNNL_TE}, it is shown that 15 seconds latency is allowable for user bids to reach the distribution level aggregator in a transactive energy scenario. Real time price signalling will also have low latency requirements.

\subsection{Data Transportation Issues}\label{subsec:datadddd}
Since data collected from different points in the NAN are used to gain awareness of the system state, any unintentional changes to the data during transportation will have an adverse impact on the overall system. In \cite{4D}, the authors identify issues that might affect data as drop, delay, disorder and distortion. Dropped data refers to messages being lost in transit, delay is associated with transportation time and is a function of the network topology as well as device capabilities. When the integrity of messages is compromised, data is said to be distorted. Data gets disordered when messages reach their destination and are processed before older messages. Such data might lead to an improper sequence of control operations, thereby creating unsafe conditions.

\subsection{Challenges in NAN Implementation}
The heterogeneous characteristics of NAN communications imposes some unique challenges. Some of these issues are described below. Other areas of concern include data security and timeliness management.

\subsubsection{Optimal gateway placement} Gateways nodes form the interface between NAN and WAN. To provide coverage to a large geographical area and maintain redundancy, multiple gateways must be deployed, but the larger the number of gateways, higher will be the infrastructure cost. Several researchers have addressed the problem of minimizing the number of gateways required to service a network subject to some constraints (like upper bound of hop count, number of nodes serviced by a gateway, gateway throughput etc)\cite{gateway1,gateway2}. 
Further research opportunities exist in designing constraints that comprehensively address network requirements, and deployment algorithms that combine heuristics along with the geometric properties of the network area.

\subsubsection{Application of cognitive radio technologies} Cognitive radios can sense `spectrum holes' i.e.  the unused portions of the licensed frequency band. Since NANs operate in the unlicensed spectrum which maybe in contention with other unlicensed systems, effectively and reliably utilizing cognitive radios can increase throughput and reduce delay in NAN communications \cite{cognitive}. 

\subsubsection{Routing and scheduling algorithms} Routing and scheduling for SG information need to account for the mesh topology structure of NANs. Communications maybe unicast or multicast. Some routing solutions in literature seek to build upon the existing RPL (Routing Protocol for Low power lossy networks) while others modify the IEEE 802.11s based HWMP. Both of these protocols represent the communication network as trees. In RPL, the network topology is abstracted as destination-oriented directed acyclic graphs with no closed loops. HWMP supports static WMNs in SG NAN and trees are formed by periodic announcement of root nodes. 


\section{Problem Formulation}\label{sec:problem}
\subsection{Communication Network Model}\label{model}

The SG NAN can be considered to be a static wireless network with configurable link activation states dictated by a centralized network controller. Without loss of generality, the network topology is described as an undirected unweighted connected graph $\mathcal{G}(\mathcal{N},\mathcal{L})$ where $\mathcal{N}$ is the set of all nodes in the network and $\mathcal{L}$ is the set of all edges or links representing connectivity of the nodes. The number of nodes and links are given by $N$ and $L$ respectively. Set $\mathcal{N}$ may be partitioned into three non-overlapping subsets- $\mathcal{N_D}$, $\mathcal{N_S}$ and $\mathcal{N_R}$. The set of source nodes in the network (in SG NANs, these maybe smart meters, Intelligent Electronic Devices (IED), Home Energy Management Systems (HEMS) or HAN gateways) is denoted by $\mathcal{N_S}$. The gateway or destination nodes form set $\mathcal{N_D}$ while relay nodes form set  $\mathcal{N_R}$. Relay nodes do not inject new messages into the network and are only used to connect distant source nodes that would otherwise be disjoint.
The network operates in slotted time with slots normalized to integral units, so that slot boundaries occur at times $t \in \{0,1,2,\dots T\}$. Slot $t$ refers to the time interval $ [t, t + 1)$. 
For simplicity, we have not differentiated among links and the transmission rate of all links is considered to be a single message per time slot. 
It is assumed that a node can communicate with its neighboring nodes, but cannot operate uplink and dowlink simultaneously.

Flow of network messages is described by queues. It is assumed that nodes in $\mathcal{N_S}$ have a constant message injection rate, which is known. This is a reasonable assumption, because under normal operating conditions, rates in which periodic messages appear in the NAN do not change much. Using this assumption, the optimum link schedule is generated (described in section \ref{sec:MILP}). Since injection rates do not vary widely, the generated schedule  may be repeated. Hence, an infinite time horizon problem is converted to a finite time problem. When nodes inject new messages, their queue sizes grow. At any time slot, a node can service i.e. receive or send only one message. The queue size at nodes in $\mathcal{N_D}$ is always considered to be zero i.e. any message that reaches the destination is considered to be serviced. 
This paper considers an unconstrained routing scenario for upstream communication (from $\mathcal{N_S}$ to $\mathcal{N_D}$) i.e. messages do not have a predefined path to destination. Nodes start transmitting if one of their adjacent links is activated. Hence, nodes in $\mathcal{N_S}$ do not have to maintain tree tables containing routes to destination nodes.  In this network model, only nodes in $\mathcal{N_D}$ need to maintain predefined routes for sending unicast information downstream (from $\mathcal{N_D}$ to $\mathcal{N_S}$). This enhances the security of the network against routing table poisoning cyber attacks. 

\subsection{MILP Formulation}\label{sec:MILP}
This paper formulates an MILP to minimize the time taken by all messages in a network to reach their destination. Delay of the slowest message is minimized. The MILP solution generates an optimal schedule that determines which links are active at any time slot. If a node senses that one of its adjacent links is active and its queue is non empty, then it can transmit messages on that link. It must be noted here, that our formulation applies to periodic data. We further assume that all nodes in $\mathcal{N_D}$ send information to the same utility control center, and hence messages from the source nodes may be sent to any of  the destination nodes in  $\mathcal{N_D}$.  In this paper, vectors and matrices are denoted in bold text.

\textit{Remark 1:} If there are $N_D$ number of destination nodes in a network, and the number of messages at time $t=0$ is $m$, then the time taken by all messages to reach the destination can not be lower than  $\frac{m}{N_D}$ time slots. This holds true because the transmission rate of all links in the network has been assumed to be one message per slot and more than one node cannot transmit to any destination node at the same time.
\subsubsection{Evolution of queues at nodes}
Let $T_{max}$ be the maximum allowable delay for all messages to reach the destination. $\mathbf{Q} \in \mathds{Z}_+^{N\times T_{max}}$ describes the queue size at different nodes at different time slots. Rows denote nodes while columns denote time slots. The initial queue sizes at different nodes for $t=0$ are described by the following constraints.
\begin{align*}
&& Q_{i,t}&=0 && \forall t,
    \forall i\in \mathcal{N_D}\tag{3.1}\label{constr:Qdestination}\\
    && Q_{i,1}&=I_{R_i} && \forall i\in \mathcal{N_S}\tag{3.2}\label{constr:Qend}\\
    && Q_{i,1}&=0 && \forall i\in \mathcal{N_R}\tag{3.3}\label{constr:Qrelay}
\end{align*}
Queue size at nodes in $\mathcal{N_D}$ is always zero (constraint \eqref{constr:Qdestination}). Let the node with the slowest message injection rate inject one message into the network in $\tau$ time slots. Then, number of messages injected by other nodes in $\mathcal{N_S}$ in $\tau$ time slots is given by $\mathbf{I_R}\in \mathds{Z}_+^{|\mathcal{N_S}|} $ (constraint \eqref{constr:Qend}). Here, $\tau$ is arbitrary and $\tau \leq T_{max}$. It is assumed that messages appear at nodes at the beginning of the time horizon. Nodes in $\mathcal{N_R}$ inject zero messages (constraint \eqref{constr:Qrelay}).
Let $\mathbf{L} \in \{0,1\}^{N\times N}$ be the adjacency matrix for graph $\mathcal{G}$ where $L_{i,j}=1$, if there exists a link between nodes $i$ and $j$ and $L_{i,j}=0$, otherwise. Let us define $T_{max}$ matrices of binary control variables $\mathbf{C}^t \in\{0,1\}^{N\times N}, t \in \{0,1,\dots,T_{max}\}$. For all node pairs $(i,j)$ where $L_{i,j}=0$, $C_{i,j}^t $ is set to zero as well (constraint \eqref{constr:C0}). $C_{i,j}^t= 1$ means that link $L_{i,j}$ is active at time slot $t$.
\begin{align*}
&C_{i,j}^t\in \{1,0\}& &\forall t, \{(i,j) \in (\mathcal{N}\times \mathcal{N} )|L_{i,j}\neq0\}\tag{3.4}\label{constr:control_binary}\\
&C_{i,j}^t=0 & &\forall t,  \{(i,j) \in (\mathcal{N}\times \mathcal{N} )|L_{i,j}=0\}\tag{3.5}\label{constr:C0}
\vspace{0.05in}
\end{align*}

 Constraints \eqref{constr:control_binary}-\eqref{constr:Qevolution} describe the flow of network messages.
 
\begin{align*}
&Q_{i,t}&=\sum\limits_{k\in \mathcal{N}}C^{t-1}_{k,i}L_{k,i}+Q_{i,t-1}-\sum \limits_{j \in \mathcal{N}} C^{t-1}_{i,j}L_{i,j} 
   \\ &&\forall i,j,k\in \mathcal{N}, \forall t\geq1\tag{3.6}\label{constr:Qevolution}
\end{align*}
Constraint (\ref{constr:Qevolution}) states that the queue size at any node at time $t$ is the sum of incoming messages and its queue size minus number of outgoing messages at $(t-1)$. This automatically takes care of the integer property of $\mathbf{Q}$ and an explicit restriction need not be imposed.

\subsubsection{Network constraints} Network constraints are described by the following:
\begin{align*}
    &&  \Sigma C^{t}_{k,i}&+\Sigma C^{t}_{i,j} \leq 1 \nonumber&\forall i,j,k\in \mathcal{N}, \forall t\geq1 \tag{3.7}\label{constr:neighbornode}\\
    && C^t_{i,j}&+C^t_{j,i}\leq 1 
    &\forall i ,j \in \mathcal{N}, \forall t  \tag{3.8}\label{constr:halfduplex}\\
    && Q_{i,t} &\leq q^{max}_{i} &\forall i \in \mathcal{N}, \forall t\tag{3.9}\label{constr:Qbound}
\end{align*}

Equation (\ref{constr:neighbornode}) says that only one link connected to a node is active at a time. Equation (\ref{constr:halfduplex}) states that uplink and downlink cannot operate at a time. Equation (\ref{constr:Qbound}) imposes an optional upper bound on the queue size at any node. $\mathbf{q^{max}}$ is a vector.

 \subsubsection{Objective function}
 The objective function is formulated considering two sub-objectives- a) minimize total message delivery time; b) minimize number of undelivered messages if all messages cannot be routed to their destination within a stipulated time. We express this as a biobjective function and show that due to the structure of the problem, this can be done without loss of optimality. Let us first describe the two sub-objectives as separate problems $\mathbf{P_1}$ and $\mathbf{P_2}$. 
\begin{align*}
\min && -\mathbf{1}^T\boldsymbol{\delta} & & \forall i\in\mathcal{N}\tag{$P_1$}\label{eq:obj}\\
\textrm{s.t.}&& \delta_t &\in \{1,0\} \nonumber & \forall t\tag{3.10}\label{constr:delta_binary}\\
    && Q_{i,t}&\geq0 &\forall i \in \mathcal{N}, \forall t\tag{3.11}\label{constr:Qpositive}\\
    && \mathbf{Q}^T\mathbf{1}&\leq (\mathbf{1}-\boldsymbol{\delta}) M\tag{3.12}\label{constr:bigM}\\
    &&\textrm{equations   } & \eqref{constr:Qdestination}-\eqref{constr:Qbound}
    \end{align*}
$\mathbf{P_1}$ is used to minimize delivery time. In equation \eqref{constr:bigM}, $M$ is an arbitrary large number. $\boldsymbol{\delta}$ is a binary vector of length $T_{max}$. The left hand side of equation (\ref{constr:bigM}) expresses the number of undelivered network messages at any time. As long as there are any undelivered messages, the value of $\delta_t$ will be zero. When there are no undelivered messages i.e. queue lengths of all nodes goes to zero, $\delta_t=1$. Therefore, maximizing $\mathbf{1}^T\boldsymbol{\delta}$ is equivalent to minimizing the message delivery time. $\mathbf{P_2}$ seeks to minimize the number of network messages at $t=T_{max}$. 
\begin{align*}
    \textrm{min}& & \sum_{i=1}^N Q_{i,T_{max}} && \tag{$P_2$}\label{eq:P2}\\
    \textrm{s.t. } && \eqref{constr:Qdestination}-\eqref{constr:Qbound},\eqref{constr:Qpositive}
\end{align*}

Let us define $\mathbf{P_3}$ as follows. The optimal $\mathbf{C}^t$ matrices obtained by solving $\mathbf{P_3}$ gives the optimal link schedule.
\begin{align*}
    \min && -\mathbf{1}^T\boldsymbol{\delta} &+ \sum_{i=1}^N Q_{i,T_{max}} \tag{$P_3$}\label{eq:P3}\\
    \textrm{s.t. } & & \eqref{constr:Qdestination}&-\eqref{constr:bigM}
\end{align*}

\begin{lemma}
The minimizer of $\mathbf{P_3}$ minimizes $\mathbf{P_1}$ and $\mathbf{P_2}$.
\end{lemma}

\textit{Proof:} Let $\boldsymbol{\delta^{*}}$ minimize $\mathbf{P_3}$. Consider the following cases. 

\textit{Case I: All messages can be sent to destination within $T_{max}$.}

If there exists a solution for all messages to reach the destination within $T_{max}$, then the value of  $\mathbf{P_2}$ will be zero, since there are no undelivered messages left in the network. This is the minimum possible value of $\mathbf{P_2}$, since all its elements must be non-negative (equation \eqref{constr:Qpositive}). Thus, the value of $\mathbf{P_2}$ becomes fixed and minimizing $\mathbf{P_3}$ will be equivalent to minimizing $\mathbf{P_1}$.

\textit{Case II: All messages cannot be delivered within $T_{max}$.}

The value of $\mathbf{1^T}\boldsymbol{\delta^*}=0$, only if there exists no feasible solution for all messages to reach their destination within $T_{max}$. The value of $\mathbf{P_2}$, would however be greater than zero, since undelivered messages will be  left in the network.The value of $\mathbf{P_1}$ becomes fixed and minimizing $\mathbf{P_3}$ will be equivalent to minimizing $\mathbf{P_2}$.  Hence, we see that the biobjective formulation does not lead to loss of optimality.

This formulation may also be extended to include special scenarios. For example, a queue flooding cyber attack (Denial of Service) can be described with the following constraint. 
\begin{align}
   &&  Q_{i,t}=Q^{max}_i && \forall i\in \mathcal{N_A}, \forall t \tag{3.16}
\end{align}
Here, $\mathcal{N_A}$ represents the set of compromised nodes. Unequal transmission rates for different links can be incorporated by changing edge weights in the adjacency matrix. 

\section{Simulation and Results}\label{sec:experiments}
\begin{figure}[!]
    \centering
    \hspace*{.8in}
    \includegraphics[clip, trim=80 130 180 210, width=0.45\textwidth]{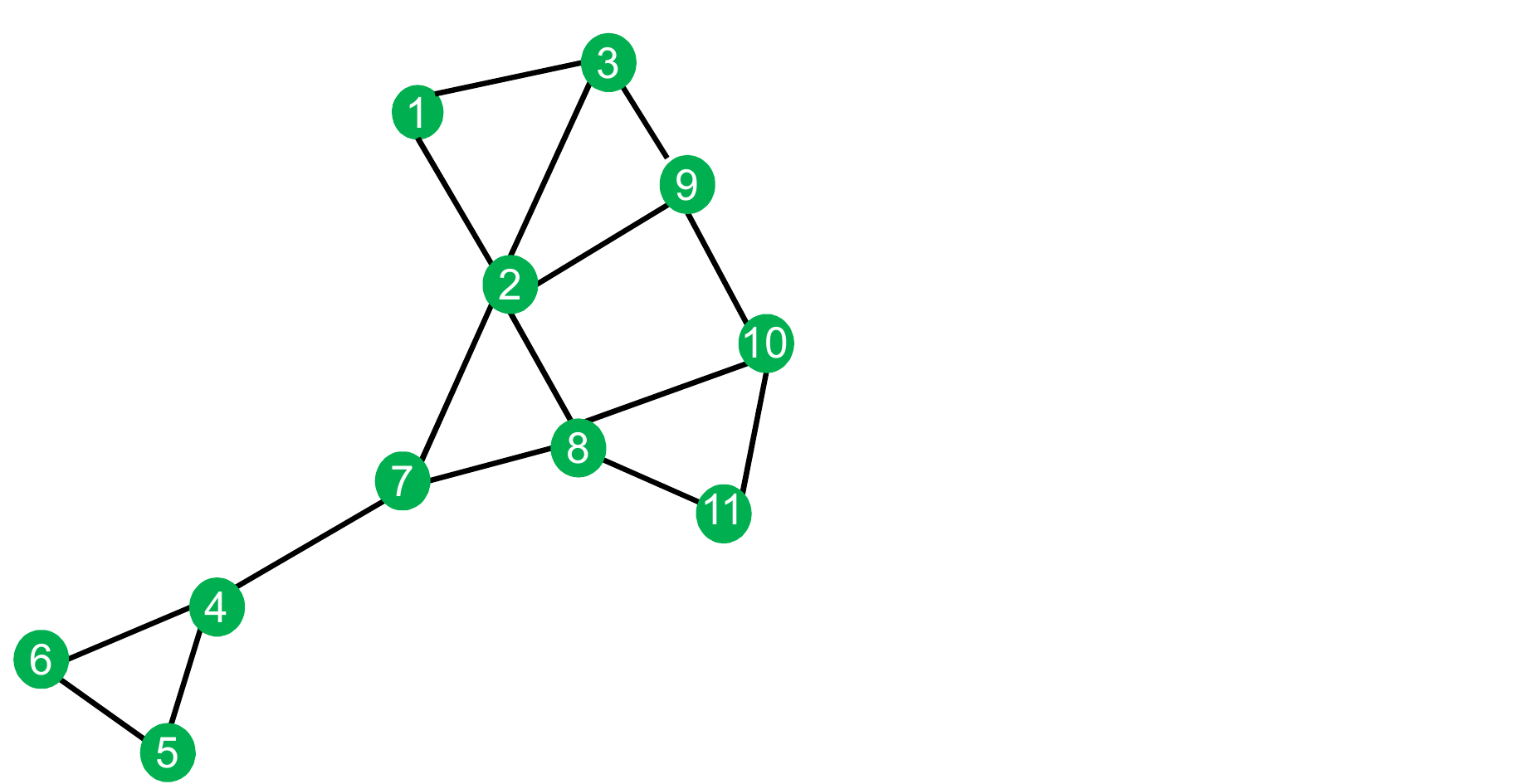}
    \caption{Network topology}
    \vspace{-0.25in}
    \label{fig:11nodegraph}
\end{figure}

This section shows how $\mathbf{P_3}$ can be used to generate optimal link schedules for a network. Results are presented for the 11-node network topology shown in figure \ref{fig:11nodegraph}. IBM CPLEX \cite{cplex} is used along with YALMIP \cite{yalmip} on MATLAB to solve the optimization problem. All computations were performed on a 9 GB RAM, Intel(R) Core$^{TM}$ i7 2.67 GHz processor machine.

\subsection{Case I: Single Gateway}
Consider the network in figure \ref{fig:11nodegraph}. Let node 1 be the only destination node. All other network nodes are considered to be sources. Relay nodes are not considered. 

\textit{\textbf{Experiment 1:} Unequal message injection rates at nodes, no upper bound on queue size}\\
 Initial queue sizes at different nodes is given by equation \eqref{eq:messagerate}. Total number of network messages is 24. The optimal link schedule generated is given by table \ref{table:opt_schedule1}. All messages reach the destination in 24 time slots, which is the theoretical lower bound of delivery time attainable as outlined in Remark 1. For ten runs, the average time taken by YALMIP and CPLEX were 0.1513 and 0.2766 seconds respectively. 
 \begin{equation}\label{eq:messagerate}
     \mathbf{I_R}^T=\left[ {\begin{array}{cccccccccc}
     1&
     3&
     3&
     2&
     3&
     1&
     2&
     3&
     3&
     3
  \end{array} } \right]\tag{4.1}
 \end{equation}
 Since no upper bound was imposed on queue sizes, the maximum queue sizes achieved at any node was 6 at nodes 4 (at $t=3,6$) and 10 (at $t=3,5$).  
 
 \textit{\textbf{Experiment 2:} Imposing an upper bound on queue size}\\
 The same problem as experiment 1 is solved with an added constraint bounding node queue sizes at three ($||\mathbf{I_R}||_\infty$). Number of time slots required is still 24. However the number of transmission operations required is 88, which is higher than the 82 operations required for experiment 1. For 10 runs, average YALMIP and CPLEX times were 0.1472 and 0.5758 seconds respectively. Optimal schedule is shown in table I. 
 \begin{table}[t!]\label{table:opt_schedule1}
     \centering
     \caption{Optimal Link Schedule}
     \begin{tabular}{p{0.4cm}|p{3.7cm}|p{3.7cm}}
        \hline \textbf{Slot} &\multicolumn{2}{|c}{\textbf{Active Links}} \\
        \hline & \textbf{Experiment 1} & \textbf{Experiment 2}\\
        \hline 0 & \{3,1\},\{7,2\},\{6,4\},\{11,10\} & \{3,1\},\{8,2\},\{4,7\}\\
        \hline 1 &\{2,1\},\{9,3\},\{5,4\},\{11,10\} & \{2,1\},\{9,3\},\{4,7\}\\
         \hline 2 & \{2,1\},\{9,3\},\{6,4\},\{11,10\} & \{3,1\},\{7,2\},\{5,4\},\{10,9\}\\
         \hline 3 &\{3,1\},\{8,2\},\{5,6\},\{4,7\},\{10,9\} & \{2,1\},\{9,3\},\{6,5\},\{4,7\},\{11,10\}\\
         \hline 4 & \{2,1\},\{9,3\},\{8,10\}& \{3,1\},\{7,2\},\{4,6\},\{10,9\},\{8,11\}\\
          \hline 5 & \{2,1\},\{7,2\},\{6,4\},\{10,8\}& \{2,1\},\{9,3\},\{6,4\},\{11,10\}\\
         \hline 6 & \{2,1\},\{9,3\},\{6,5\},\{4,7\} & \{3,1\},\{6,4\},\{10,9\}\\
          \hline 7 & \{3,1\},\{8,2\},\{10,11\},\{4,7\} & \{2,1\},\{9,3\},\{5,6\},\{4,7\},\{10,11\} \\
         \hline 8 & \{3,1\},\{7,2\},\{10,9\}& \{3,1\},\{7,2\},\{4,6\},\{11,8\},\{10,9\}\\
         \hline 9 &\{3,1\},\{7,2\},\{10,9\},\{11,8\}& \{3,1\},\{9,2\},\{5,4\},\{8,7\},\{11,10\}\\
         \hline 10 &\{2,1\},\{9,3\},\{10,8\},\{4,7\} &\{2,1\},\{9,3\},\{7,8\},\{11,10\} \\
         \hline 11 & \{3,1\},\{7,2\},\{4,6\},\{8,10\}& \{2,1\},\{9,3\},\{4,7\},\{8,10\}\\
          \hline 12 & \{2,1\},\{10,9\},\{5,6\},\{4,7\},\{8,11\}& \{3,1\},\{7,2\},\{6,5\},\{10,9\}\\
         \hline 13 & \{3,1\},\{4,7\},\{11,8\}& \{3,1\},\{7,2\},\{5,6\}\{10,9\}\\
         \hline 14 & \{2,1\},\{9,3\},\{6,4\},\{7,8\}& \{2,1\},\{9,3\},\{6,4\},\{7,8\}\\
          \hline 15 & \{2,1\},\{9,3\},\{6,5\},\{4,7\},\{8,10\}& \{3,1\},\{8,2\},\{4,7\},\{10,9\}\\
         \hline 16 & \{3,1\},\{8,2\},\{5,4\},\{10,9\} & \{2,1\},\{9,3\},\{7,8\}\\
         \hline 17 & \{2,1\},\{9,3\},\{4,7\}& \{3,1\},\{6,4\},\{8,10\}\\
         \hline 18 & \{3,1\},\{7,2\},\{10,9\}& \{2,1\},\{9,3\},\{4,7\} \\
         \hline 19 & \{2,1\},\{9,3\} & \{3,1\},\{6,4\},\{7,8\}\\
          \hline 20 & \{3,1\},\{7,2\}& \{3,1\},\{8,2\},\{4,7\},\{10,9\} \\
         \hline 21 & \{2,1\}& \{2,1\},\{9,3\},\{7,8\}\\
         \hline 22 & \{3,1\},\{7,2\}& \{3,1\},\{8,2\}\\
         \hline 23 &\{2,1\}&\{2,1\}\\
         \hline
     \end{tabular}
 \vspace{-0.1in}
 \end{table}

\textit{\textbf{Experiment 3:} Adding a relay node}\\
In the network of figure \ref{fig:11nodegraph}, node 7 is designated as a relay node that connects nodes 4, 5 and 6 to the rest of the network. The MILP solution generates a schedule to route all messages to the destination in 23 time slots. The schedule is not included due to space constraints. For 10 runs, average YALMIP and CPLEX time were 0.1573 and 0.37 seconds respectively.

\textit{\textbf{Experiment 4:} $T_{max}<\frac{m}{N_D}$\\}
In experiment 1, $T_{max}$ is limited at 20 slots, at the end of which 5 network messages are left undelivered. The optimal schedule generated matches the result for experiment 1 upto 20 slots. Average YALMIP and CPLEX time for 10 runs were respectively 0.1494 and 0.325 seconds.

\textit{\textbf{Experiment 5:} Example Application: Bid aggregation in a Transactive Energy (TE) Market}
\begin{figure}[t!]
    \centering
    \includegraphics[clip, trim=15 5 8 30,width=\columnwidth]{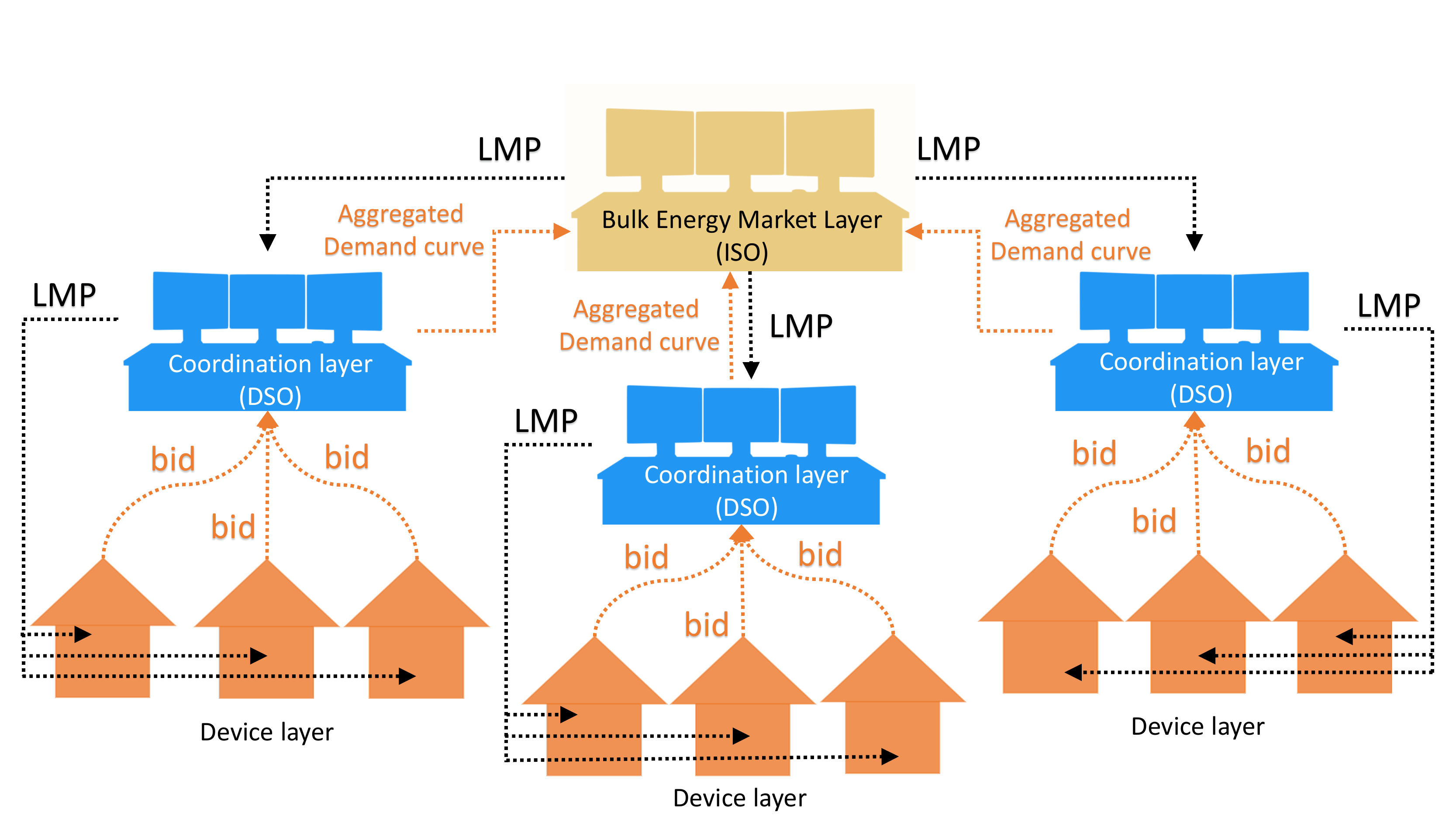}
    \caption{Hierarchical Transactive Energy Market Structure}
    \label{fig:TE}
    \vspace{-0.2in}
\end{figure}

Consider the hierarchical TE framework shown in figure \ref{fig:TE}. The control strategy for such a system is elaborated in \cite{PNNL_TE,my_thesis}. In the TE market, a central decision making authority decides the LMP (Locational Marginal Price) for the next market interval ($\sim5$ minutes) and communicates it to the market participants with responsive loads. In a distribution grid, these could be households with controllable HVAC. In response, loads modify their consumption. They also formulate a bid expressing their need for energy that is communicated upstream and used to decide LMP for the next period. \cite{PNNL_TE} shows that 15 seconds is an acceptable delay for user generated bids to reach the Distribution System Operator (DSO). 

We assume that all bids are generated at the same time by all participating consumers and the packet sizes are similar. It is important to route all bids to the DSO in the shortest time possible because the LMP calculation for the next market period depends on these bids. 
The bid aggregation problem for the network in figure \ref{fig:11nodegraph} is solved using the proposed MILP approach. A single gateway is considered at node 1. The initial queue size at all other nodes is considered 1. An optimal link schedule (table II) is generated to route all bids to the destination in 10 time slots. Average YALMIP  and CPLEX time for 10 runs were 0.1156 and 0.1158 seconds respectively.

The experiment was repeated for a random network of 100 nodes with 1 gateway node and initial queue size at all nodes set at 1. The optimal schedule routed all messages to the gateway within 100 time slots. Average YALMIP and CPLEX time for 10 runs were 1.56 and 54.34 seconds respectively.

\begin{table}\label{table:opt_schedule5}
     \centering
     \caption{Optimal Link Schedule for Bid Aggregation}
     \begin{tabular}{c|c}
        \hline \textbf{Slot}& \textbf{Active Links}\\
        \hline 0 & \{3,1\},\{7,2\},\{6,4\},\{10,9\} \\
        \hline 1 &\{2,1\},\{9,3\},\{4,7\},\{11,10\} \\
         \hline 2 & \{2,1\},\{9,3\},\{5,4\},\{8,10\} \\
         \hline 3 &\{3,1\},\{7,2\},\{10,9\} \\
         \hline 4 & \{3,1\},\{9,2\},\{4,7\}\\
          \hline 5 & \{2,1\},\{7,8\},\{10,9\}\\
         \hline 6 & \{2,1\},\{9,3\},\{8,10\},\{4,7\}\\
          \hline 7 & \{3,1\},\{7,2\},\{10,9\} \\
         \hline 8 & \{2,1\},\{9,3\}\\
         \hline 9 &\{3,1\}\\
         \hline
     \end{tabular}
     \vspace{-0.2in}
 \end{table}

\subsection{Case II: Multiple Gateways}
Experiment 5 was repeated with 2 gateway nodes. The first gateway was placed at node 1 and the position of the second one was varied. Performance under different gateway locations is shown in table III. It is evident that gateway placement has an impact on delivery time and throughput. Optimal placement of gateway nodes is not investigated in this paper.
\begin{table}[t]\label{table:multiplegateway}
     \centering
     \caption{Second Gateway Node placed at Different Locations}
     \begin{tabular}{p{1.5cm}|p{1.45cm}|p{2cm}|p{2.05cm}}
        \hline \textbf{Second Gateway}& \textbf{Delivery Time (slots)}&\textbf{Messages Sent to Node 1}&\textbf{Messages Sent to Second Gateway}\\
        \hline 2 & 7 & 2 &7\\
        \hline 3 & 9 &5 &4  \\
         \hline 4 & 5 &5 &4\\
         \hline 5& 5 &5 &4\\
         \hline 6 & 5 &5 &4\\
          \hline 7 & 5 &4 &5\\
         \hline 8 & 7 &4 &5\\
          \hline 9 & 8 &4 &5\\
         \hline 10 & 8 &5 &4\\
         \hline 11 & 8 &4 &5\\
         \hline
     \end{tabular}
     \vspace{-0.2in}
 \end{table}
\section{Limitations and Future Directions}\label{sec:extension}
The proposed approach is not without its limitations and further research is required to adequately address the question of optimal message routing and link scheduling in SG NANs. First, the proposed method is static or proactive and needs to be extended to address the question of dynamic routing in case of link failures and other topological changes. Moreover, the proposed approach needs to be extended to include asymmetric link capacities and unequal message sizes. Also, since MILPs are NP-hard problems, devising heuristic algorithms to devise schedules for larger networks could be beneficial.
Differentiating messages on the basis of priority is another direction that the authors want to pursue.
\section{Conclusion}\label{sec:conclusion}
A MILP based formulation for generating an optimal link schedule for communication in SG NANs is proposed in this paper. This strategy increases resiliency against routing table poisoning attacks and is shown to work in a number of different network conditions  with different constraints. The scheme is also applied to find an optimal schedule for aggregating user generated bids in a transactive energy market. Although the MILP is NP-hard, it is shown that a solution can be found within reasonable time for a network with 100 nodes. Limitations of the proposed approach and future directions for further research are also outlined.  

\section*{Acknowledgment}
The authors would like to thank Chiranjib Saha and Manish Kumar Singh from Virginia Tech for their valuable inputs that helped shape this paper.

\bibliographystyle{IEEEtran}
\bibliography{bibliography.bib}

\vspace{12pt}
\end{document}